\documentclass[12pt]{article}
\usepackage[dvips]{color}
\usepackage{amsmath}
\usepackage{graphicx}
\usepackage{subfigure}
\usepackage{epsfig}
\usepackage{amsmath}
\usepackage{cite}
\usepackage{color}
\usepackage{subfigure}

\begin{document}
\begin{center}
\Large{\bf Tsallis holographic dark energy under Complex form of Quintessence model}\\
\small \vspace{1cm} {\bf J. Sadeghi$^{\star}$\footnote {Email:~~~pouriya@ipm.ir}}, \quad
{\bf S. Noori Gashti$^{\star}$\footnote {Email:~~~saeed.noorigashti@stu.umz.ac.ir}}, \quad and
{\bf T. Azizi$^{\star}$\footnote {Email:~~~t.azizi@umz.ac.ir}}, \quad
\\
\vspace{0.5cm}$^{\star}${Department of Physics, Faculty of Basic
Sciences,\\
University of Mazandaran
P. O. Box 47416-95447, Babolsar, Iran}\\
\small \vspace{1cm}
\end{center}
\begin{abstract}
In this paper, we use a Tsallis holographic dark energy model in two forms, interacting and non-interacting cases, to acquire some parameters as the equation of state for the energy density of the Tsallis model in the FRW universe concerning the complex form of quintessence model. We will study the cosmology of complex quintessence by revamping the potential and investigating the scalar field dynamics. Then we analyze ($\omega-\omega'$) and stability in two cases, i.e., non-interacting and interacting. We will explore whether these cases describe a real universe by calculating fractional energy density $\Omega_{D}$ and concerning two parts of the quintessence field effect ( complex and real part ) by considering the real part of this field to be a slow-roll field.
We know that the part in which the fractional energy density ($\Omega_{D} > 1$) does not describe a real universe. Also, we specified an interacting coupling parameter $b^{2}$ that depends on the constant parameter of the Tsallis holographic model ($\delta$) with respect to fractional energy density ($0.73$). Unlike independence between the fractional energy density and interacting coupling in the real quintessence model, we determine a relationship among these parameters in this theory. Finally, by plotting some figures, we specify the features of ($\omega-\omega'$) and ($\nu_{s}^{2}$) in two cases and compare the result with each other.
\end{abstract}
\newpage
\tableofcontents

\section{Introduction}
One of the essential issues facing cosmologists is explaining how the universe expanded. The universe's accelerated expansion has been proven by various observations such as CMB anisotropies, supernovae Ia, and large-scale structure\cite{1,2,3}.
So far, cosmologists have introduced various theories to determine the universe's accelerating expansion and have compared the results of these theories with the latest observable data. Among the theories proposed, we can name inflation and the dark energy with unknown nature, which with negative pressure, has led to the universe's accelerated expansion and is one of the most accepted theories in discussing the universe's accelerated expansion.
Cosmologists have introduced different structures for such a theory, and the results of these models have been evaluated. The cosmological constant is the most specific model for analyzing the nature of dark energy structure\cite{7,8}.
Of course, according to all these explanations, we always face problems, the most important of which can be called fine-tuning problems.
Among the most important models that cosmologists have introduced to study the nature of dark energy
can be named interacting dark energy models, braneworld models, Chaplygin gas models, phantom, ghost condensate, quintom quintessence, K-essence, tachyon models \cite{9,9b,9c,9d,10,11,11b,12,13,14,15,15b,16,17,18}.\\
An important point in studying the structure of dark energy is introducing a more significant number of degrees of freedom than standard cosmology. one should research these degrees of greater freedom, properties, and consequences in modeling the universe.
Another example is the ghost dark energy that uses the Veneziano ghost to explain the universe's expansion, which has also recently been worked on\cite{19,20,20b}.
In this model, the cosmological constant considered arises from the contribution of ghost fields. Ghosts were introduced to resolve the U (1) problem\cite{21,22,23,24,25}.
 Among other models introduced to study the concepts is called holographic dark energy.
 In this regard,  researchers studied various structures and compared the results with other models and the latest observable data\cite{26,27,28}.\\
There is also a new entropy in generalized statistical mechanics for black holes that is different from the Bekenstein entropy and has led to the introduction of a new holographic dark energy model called Tsallis holographic dark energy\cite{29,30,31,32,33,34,35,36,37}.
Also, other models, such as Kaniadakis Holographic Dark Energy, etc., are introduced by generalizing this entropy. Each of these models is examined in different structures and conditions. The results are evaluated and compared with the latest observable data\cite{38,39,40}.\\
As mentioned, one of the most famous models in describing the nature of dark energy is a scalar field theory called quintessence, which represents a scalar field with a parameter (Q) and a decreasing potential.
In describing dark energy with negative pressure, it is interpreted that if the field evolves slowly, the potential energy density is greater than the kinetic energy density, which leads to negative pressure, meaning that the universe follows an accelerated expansion.
The quintessence field has been investigated and analyzed in two forms.
In \cite{41,42}, They examined the basic features of real quintessence theory.
 Another example in the form of a complex scalar field has been studied to describe the accelerated expansion of the universe \cite{43,44,45}.
This model has also been used to describe dark energy using correspondence between ghost dark energy and the complex quintessence \cite{46}.
So far, the study of dark energy from a holographic perspective concerning the structure of complex quintessence has not been studied. This article examines the correspondence between the Tsallis holographic dark energy and complex quintessence from two perspectives (interacting and non-interacting cases).\\
First, we review the fundamental equations of the two theories viz  Tsallis holographic dark energy (THDE) and complex quintessence field (CQF) and propose a correspondence between the two scenarios.
Then we analyze ($\omega-\omega'$) and stability in two cases, i.e., non-interacting and interacting. We will explore whether these cases describe a real universe by calculating fractional energy density $\Omega_{D}$ and concerning two parts of the quintessence field effect ( complex and real part ) by considering the real part of this field to be a slow-roll field.
We know that the part in which the fractional energy density ($\Omega_{D} > 1$) does not describe a real universe. Also, we specified an interacting coupling parameter $b^{2}$ that depends on the constant parameter of the Tsallis holographic model ($\delta$) concerning fractional energy density ($0.73$). Unlike independence between the fractional energy density and interacting coupling in the real quintessence model, we determine a relationship among these parameters in this theory. Finally, by plotting some figures, we specify the features of ($\omega-\omega'$) and ($\nu_{s}^{2}$) in two cases and compare the result with each other.

\section{Complex form of Quintessence Field}

This section will first introduce and express the basic equations for the Complex Quintessence Field and detail the results. Hence the Friedmann-Robertson-Walker metric is described in the following form\cite{45},

\begin{equation}\label{1}
\textrm{d}s^{2}=-\textrm{d}t^{2}+a^{2}(t)\big(\frac{\textrm{d}r^{2}}{1-kr^{2}}+r^{2}\textrm{d}\Omega^{2}\big),
\end{equation}

\hspace{-0.6cm}$k$ represents the space curvature for the flat, open, and closed universe is 0, -1, 1, respectively. The action of this universe is as follows form,

\begin{equation}\label{2}
S=\int \textrm{d}^{4}x\sqrt{-g}\big(\frac{1}{16\pi G}R+\rho_{m}+\mathcal{L}_{\Phi}\big),
\end{equation}

\hspace{-0.6cm}where $g$, $G$, $R$, and $\rho_{m}$ are determinants of the metric tensor $g_{\mu\nu}$, the Newton’s constant, the Ricci scalar, and the density of ordinary matter, respectively. We also define the Lagrangian density of the complex form of the quintessence field.

\begin{equation}\label{3}
\mathcal{L}_{\Phi}=\frac{1}{2}g^{\mu\nu}(\partial_{\mu}\Phi^{\star})(\partial_{\nu}\Phi)-V(|\Phi|),
\end{equation}

\hspace{-0.6cm}where $\mu,\nu=0,1,2,3$. As shown in the above equation, we hypothesized that the potential V depends only on the absolute values of the complex form of the quintessence scalar field. Now we want to use alternative fields in these equations; we express the complex quintessence scalar field in terms of amplitude $\phi$ and the phase $\theta$ in the following form.

\begin{equation}\label{4}
\Phi(x)=\phi(x)\textrm{e}^{i\theta(x)}.
\end{equation}

A more accurate equation (4) can be expressed as $\Phi(t)=\phi(t)\textrm{e}^{i\theta(t)}$. The use of newly defined variables ($\phi(x)$) and ($\theta(x)$) helps to calculate the reconstructed equations, which will lead to the relationship between SNe Ia data and quintessence potential. Therefore, by using equation (4), the Lagrangian density for the new variable, which is given by,

\begin{equation}\label{5}
\mathcal{L}_{\Phi}=\frac{1}{2}g^{\mu\nu}(\partial_{\mu}\phi)(\partial_{\nu}\phi)+\frac{1}{2}\phi^{2}g^{\mu\nu}(\partial_{\mu}\theta)(\partial_{\nu}\theta)-V(\phi).
\end{equation}

A variation of the action (equation (2)) with the Lagrangian density obtained in the above equation will lead to calculating Einstein equations and field equations of the complex form of the quintessence scalar field. So using the metric tensor (equation (1)), we will have.

\begin{equation}\label{6}
H^{2}\equiv(\frac{\dot{a}}{a})^{2}=\frac{8\pi G}{3}\rho-\frac{k}{a^{2}}=\frac{8\pi G}{3}\big(\rho_{m}+\frac{1}{2}(\dot{\phi}^{2}+\phi^{2}\dot{\theta}^{2})+V(\phi)\big)-\frac{k}{a^{2}},
\end{equation}

\begin{equation}\label{7}
(\frac{\ddot{a}}{a})^{2}=-\frac{4\pi G}{3}(\rho+3p)=-\frac{8\pi G}{3}\big(\frac{1}{2}\rho_{m}+(\dot{\phi}^{2}+\phi^{2}\dot{\theta}^{2})-V(\phi)\big),
\end{equation}

\begin{equation}\label{8}
\ddot{\phi}+3H\dot{\phi}-\dot{\theta}^{2}\phi+V'(\phi)=0,
\end{equation}

\begin{equation}\label{9}
\ddot{\theta}+(2\frac{\dot{\phi}}{\phi}+3H)\dot{\theta}=0.
\end{equation}

In the above equations, $H$, $dot$, and $'$ show the Hubble parameter, time derivative, and derivative concerning the parameter $\phi$, respectively. Also, $p$ and $rho$ represent the pressure and the energy density. As seen from the above equations, the two equations (6) and (7) form the Friedman equations for the model. The above equations are also fundamental equations that govern the universe's evolution. It can note that non-relativistic matter contributes energy density $\rho_{M}$ and pressure $p_{M}=0$, while the evolution of a complex scalar field helps to form energy density $\rho_{\Phi}$ and pressure $p_{\Phi}$ in the following form.

\begin{equation}\label{10}
\rho_{\Phi}=\frac{1}{2}(\dot{\phi}^{2}+\phi^{2}\dot{\theta}^{2})+V(\phi),
\end{equation}

\begin{equation}\label{11}
p_{\Phi}=\frac{1}{2}(\dot{\phi}^{2}+\phi^{2}\dot{\theta}^{2})-V(\phi).
\end{equation}

Also, We can solve equation (9) and get a solution for angular velocity, which is expressed in the following form.

\begin{equation}\label{12}
\dot{\theta}=\frac{\omega}{a^{3}\phi^{2}}.
\end{equation}

As is apparent in the above equation, the parameter $\omega$ is an integration constant determined according to the initial conditions of the parameter $\dot{\theta}$. Now, using equation (12), we can quickly rewrite equations (6-9) in terms of parameter $\phi$, which can be easily calculated. In the next section, we will explain the fundamental equations of Tsallis holographic dark energy and calculate some parameters needed to be investigated in this paper.

\section{Tsallis holographic dark energy}

We know that gravity is a long-range interaction, so we can also use the generalized structure of statistical mechanics to study gravitational systems.
We also know that the entropy of black holes can be studied by generalizing and extending the entropy of Bekenstein.
One of these entropies is the application of Tsallis statistics to the system.
Various dark energy holographic structures such as THDE, SMHDE, and RHDE have been introduced recently.
Each model has its characteristics and can be converted with direct calculations.
Meanwhile, THDE is also built using the generalized entropy of Tessalis and is not stable at the classical level.
This model has been studied with different conditions, and the results have been compared with other dark energy models and the latest observable data\cite{46,47,48,49,50,51,52}.
But in this article, we are looking to examine specific conditions using this model.
Hence, in a nonflat FRW universe containing dark matter and THDE, the Friedman equation is expressed as follows.

\begin{equation}\label{13}
H^{2}+\frac{k}{a^{2}}=\frac{1}{\widetilde{r}_{A}^{2}}=\frac{8\pi G}{3}(\rho_{m}+\rho_{D}),
\end{equation}

\hspace{-0.6cm}where $\rho_{D}$ and $\rho_{M}$ are representations of the energy density of THDE and pressureless DM, respectively. The Tsallis holographic energy density is defined as,

\begin{equation}\label{14}
\rho_{D}=BL^{2\delta-4},
\end{equation}

\hspace{-0.6cm}where $B$ is an unknown parameter; also, by assuming the Hubble horizon as the IR cutoff $L=H^{-1}$, the energy density converts the following form,

\begin{equation}\label{15}
\rho_{D}=BH^{4-2\delta}.
\end{equation}

We can introduce other energy densities,i.e., curvature and critical energy density, to calculate some fractional energy densities. So three fractional energy are as follows,

\begin{equation}\label{16}
\Omega_{D}=\frac{8\pi G \rho_{D}}{3H^{2}},\hspace{5pt}\Omega_{m}=\frac{8\pi G \rho_{m}}{3H^{2}},\hspace{5pt}\Omega_{k}=\frac{k}{H^{2}a^{2}}.
\end{equation}

Now, according to the above definitions, the Friedmann equation can be rewritten as follows

\begin{equation}\label{17}
\Omega_{m}+\Omega_{D}=1+\Omega_{k}.
\end{equation}

We now consider two different forms. If there is no interaction between matter and the Tsallis holographic dark energy, the equations are expressed as,

\begin{equation}\label{18}
\dot{\rho}_{D}+3H\rho_{D}(1+\omega_{D})=0,
\end{equation}

\begin{equation}\label{19}
\dot{\rho}_{m}+3H\rho_{m}=0.
\end{equation}

Therefore, the equation of state for Tsallis holographic dark energy is expressed in the following form for this case.

\begin{equation}\label{20}
\omega_{D}=-\frac{(\delta-2)(\Omega_{k}+3)+3}{3((\delta-2)\Omega_{D}-1)}.
\end{equation}

Consider the second case as an interaction between matter and Tsallis holographic dark energy, in which case we will have

\begin{equation}\label{21}
\dot{\rho}_{D}+3H\rho_{D}(1+\omega_{D})=-\mathcal{Q},
\end{equation}

\begin{equation}\label{22}
\dot{\rho}_{m}+3H\rho_{m}=\mathcal{Q}.
\end{equation}

According to the above equation, the parameter $\mathcal{Q}$ is called the interaction term. This interaction parameter is expressed in the following form.

\begin{equation}\label{23}
\mathcal{Q}=3b^{2}H(\rho_{D}+\rho_{m})=3b^{2}H\rho_{D}(1+r),
\end{equation}

\hspace{-0.6cm}where $b^{2}$ is a coupling parameter and $r=\frac{\rho_{m}}{\rho_{D}}=-1+\frac{1}{\Omega_{D}}(1+\Omega_{k})$. $\omega_{D}=\frac{p_{D}}{\rho_{D}}$ specifies the equation of state, which in this case is also calculated for Tsallis holographic dark energy as follows,

\begin{equation}\label{24}
\omega_{D}=-\frac{3+(\delta-2)(\Omega_{k}+3)+3b^{2}(1+r)}{3(1+(\delta-2)\Omega_{D})}.
\end{equation}

Considering all the equations and computational values in the two previous sections, we will continue our calculations as mentioned in the text. Of course, you can see more details of the above measures about Tsallis holographic dark energy and Complex Quintessence Field in\cite{45,51,52}.

\section{THDE and CQF in FRW universe (non-interacting case) }

First, we consider the non-interaction case and create the correspondence between the energy density of the complex form of the quintessence field and the Tsallis holographic dark energy according to equations (10), (12), and (15). In that case, we have a combination of the above equations.

\begin{equation}\label{25}
\rho_{D}=\frac{1}{2}(\dot{\phi}^{2}+\frac{\omega^{2}}{a^{6}\phi^{2}})+V(\phi)=B H^{4-2\delta}.
\end{equation}

We set $\mathcal{T}=\frac{1}{2} (\dot{\phi}^{2}+\frac{\omega^{2}}{a^{6}\phi^{2}})$, so we rewrite the above equation  as

\begin{equation}\label{26}
V(\phi)=B H^{4-2\delta}-\mathcal{T}.
\end{equation}

Since the purpose is to establish a correspondence between the energy density of the complex form of quintessence field and the Tsallis holographic dark energy, we will have,

\begin{equation}\label{27}
\omega_{\Phi}\equiv\frac{p_{\Phi}}{\rho_{\Phi}}=\omega_{D}.
\end{equation}

Concerning mentioned points and equations (10),(11), and (20) ones calculate,

\begin{equation}\label{28}
\frac{\mathcal{T}-V(\phi)}{\mathcal{T}+V(\phi)}=-\frac{(\delta-2)(\Omega_{k}+3)+3}{3((\delta-2)\Omega_{D}-1)}.
\end{equation}

The potential of the above relation can easily rewrite.

\begin{equation}\label{29}
V(\phi)=-\frac{(-3\delta-6\Omega_{D}+3\delta\Omega_{D}+2\Omega_{k}-\delta\Omega_{k})}{(-2+\delta)(3+3\Omega_{D}+\Omega_{k})}\times\mathcal{T}.
\end{equation}

By combining two equations (26) and (29), one can calculate,

\begin{equation}\label{30}
B H^{4-2\delta}-\mathcal{T}=-\frac{(-3\delta-6\Omega_{D}+3\delta\Omega_{D}+2\Omega_{k}-\delta\Omega_{k})}{(-2+\delta)(3+3\Omega_{D}+\Omega_{k})}\times\mathcal{T}.
\end{equation}

With the straightforward calculation of the above equation, we will have,

\begin{equation}\label{31}
H=6^{\frac{1}{4-2\delta}}\bigg(-\frac{\mathcal{T}(-1+(-2+\delta)\Omega_{D})}{B(-2+\delta)(3+3\Omega_{D}+\Omega_{k}}\bigg)^{\frac{1}{4-2\delta}}.
\end{equation}

Here we point out an important point in this article: we consider $(k=0)$ and perform physical interpretations such as $\omega-\omega'$ and stability analysis for the non-interacting case. In the literature, $\omega-\omega'$ is an essential tool used to distinguish different models that have been frequently discussed in the literature; as mentioned in the text, $\omega$ is the equation of state, and $\omega'$ is derivative from $\omega$ concerning $\ln a$. We continue the calculations of this article.

\begin{equation}\label{32}
\frac{\textrm{d}\Omega_{D}}{\textrm{d}\ln a}=\frac{(-2+2\delta)}{3}\Omega_{D}(1+q).
\end{equation}

q is a deceleration parameter defined as

\begin{equation}\label{33}
q=-1-\frac{\dot{H}}{H^{2}}.
\end{equation}

So,

\begin{equation}\label{34}
q=\frac{1+\Omega_{k}+(1-2\delta)\Omega_{D}}{2+2(\delta-2)\Omega_{D}}.
\end{equation}

With respect to above equations, we will have,

\begin{equation}\label{35}
\frac{\textrm{d}\omega_{D}}{\textrm{d}\ln a}=-\frac{(-2+\delta)(-1+\delta)\Omega_{D}(3-3\Omega_{D}+\Omega_{k})(3+(-2+\delta)(3+\Omega_{k}))}{3(-1+(-2+\delta)\Omega_{D})^{2}(1+(-2+\delta)\Omega_{D})}.
\end{equation}

So the $\omega'$ for (k=0) which is calculated as,

\begin{equation}\label{36}
\omega'=-\frac{3(1+\delta(-1+\omega_{D})-3\omega_{D})\omega_{D}(-1+\delta+\omega_{D})}{(-2+\delta)(-1+\delta+2\omega_{D})}.
\end{equation}

The above equation is used $\omega-\omega'$ analysis for the mentioned model and non-interactive case. We will analyze the results by plotting a figure; also, there are different ways to study the stability of the model, which here uses the sound speed.

\begin{equation}\label{37}
\nu_{s}^{2}(z)\equiv\frac{\textrm{d}p_{D}}{\textrm{d}\rho_{D}}=\frac{\textrm{d}p_{D}/dz}{\textrm{d}\rho_{D}/dz},
\end{equation}

\hspace{-0.6cm}where $z$ is redshift, and we have $1+z=a^{-1}$, for the $k=0$ the hobble constat which is calculated as,

\begin{equation}\label{38}
H=\big(\frac{3M_{p}^{2}}{B(1+r)}\big)^{\frac{1}{2-2\delta}},
\end{equation}

\hspace{-0.6cm}where $M_{p}^{2}=\frac{1}{8\pi G}$. So one can calculate

\begin{equation}\label{39}
\frac{\textrm{d}H}{\textrm{d}a}=\frac{3^{-1+1/2-2\delta}(3-3\Omega_{D})(M_{p}/B\Omega_{D})^{1/2-2\delta}}{2a(1+(-2+\delta)\Omega_{D})}.
\end{equation}

The stability is specified with $\rho_{D}$ and $p_{D}$, so we will have

\begin{equation}\label{40}
\frac{\textrm{d}\rho_{D}}{\textrm{d}z}=B\frac{\textrm{d}H^{4-2\delta}}{\textrm{d}z}=B(-a^{2})\frac{\textrm{d}H^{4-2\delta}}{\textrm{d}a},
\end{equation}

\begin{equation}\label{41}
\frac{\textrm{d}p}{\textrm{d}z}=\frac{\textrm{d}\omega_{D}}{\textrm{d}z}\rho_{D}+\omega_{D}\frac{\textrm{d}\rho_{D}}{\textrm{d}z}=-a\frac{d\omega_{D}}{d\ln a}\rho_{D}+\omega_{D}\frac{\textrm{d}\rho_{D}}{\textrm{d}z}.
\end{equation}

So with respect to equations (35-41), one can obtain

\begin{equation}\label{42}
\nu_{s}^{2}(z)=\frac{(-2+\delta)(3+3\Omega_{D}+\Omega_{k})}{-3+3(-2+\delta)\Omega_{D}}.
\end{equation}

It can be a fair expectation that we want $\omega-\omega'$ and stability analysis for the complex form of the quintessence model to have similar results when the real part of this model is used.
Since these two analyses are related to Tsallis holographic dark energy and unrelated to their complex part, we wanted to identify the effects of the complex part of the quintessence field in these analyses and examine its impact.
Therefore, in the following, we will limit our calculations to the effects of the slow-rolling field and advance our computational process.
Also, after reviewing this part, we will develop an interacting case about the mentioned model and compare the results of these two parts. So concerning equations (10), (11) and the definition of $\mathcal{T}$, we calculate,

\begin{equation}\label{43}
\dot{\phi}^{2}+\frac{\omega^{2}}{a^{6}\phi^{2}}=B H^{4-2\delta}\bigg(\frac{(-2+\delta)(3+3\Omega_{D}+\Omega_{k})}{-3+3(-2+\delta)\Omega_{D}}\bigg).
\end{equation}

According to the analysis that is done in terms of slow-rolling structure. So by ignoring the term $\dot{\phi}^{2}$, we will have

\begin{equation}\label{44}
\phi=\frac{\sqrt{3}\omega H^{\delta}}{a^{3}H^{2}}\times\sqrt{\frac{2\Omega_{D}+1-\delta \Omega_{D}}{6B-3B\delta+6B\Omega_{D}-3B\delta\Omega_{D}+2B\Omega_{k}-B\delta\Omega_{k}}}.
\end{equation}

After solving, we considered only the positive part of the solution. Also, since $\dot{\phi}=H\frac{\textrm{d}\phi}{\textrm{d}\ln a}$ and can not ignore the parameter $H$, so $\frac{\textrm{d}\phi}{\textrm{d}\ln a}\approx 0$. Hence by combining equations (32), (39), and (44), for $k=0$ we will have.

\begin{equation}\label{45}
\frac{\textrm{d}\phi}{\textrm{d} \ln a}=\frac{(-1+\delta)^{2}(-1+\Omega_{D})\Omega_{D}H^{-2+\delta}\omega}{2a^{3}(1+\Omega_{D})\sqrt{-B(-2+\delta)(1+\Omega_{D})}\sqrt{1-(-2+\delta)\Omega_{D}}(1+(-2+\delta)\Omega_{D})}\approx 0.
\end{equation}

So with the above explanation, we have.

\begin{equation}\label{46}
(-1+\delta)^{2}(\Omega_{D}^{2}-\Omega_{D})=0.
\end{equation}
The solution of the above equation is 0 and 1, and we know that the parameter $\Omega_{D}$ must be a value less than 1, which is not acceptable for 1. We can say that for the unacceptable values of this solution viz $1$; we can not use the non-interacting case if we consider the complex form of quintessence field as Tsallis holographic dark energy and create the “slow-rolling” field for describing the real universe. Actually, using the complex part of the field here has an important role that can not be the acceptable solution for the universe evolution for the non-interacting case. So we will investigate the interacting case to specify the important role of the complex part of the quintessence field in universe development. The next section evaluates the interacting case and compares the resul.
\begin{figure}[h!]
 \begin{center}
 \subfigure[]{
 \includegraphics[height=5.5cm,width=5.5cm]{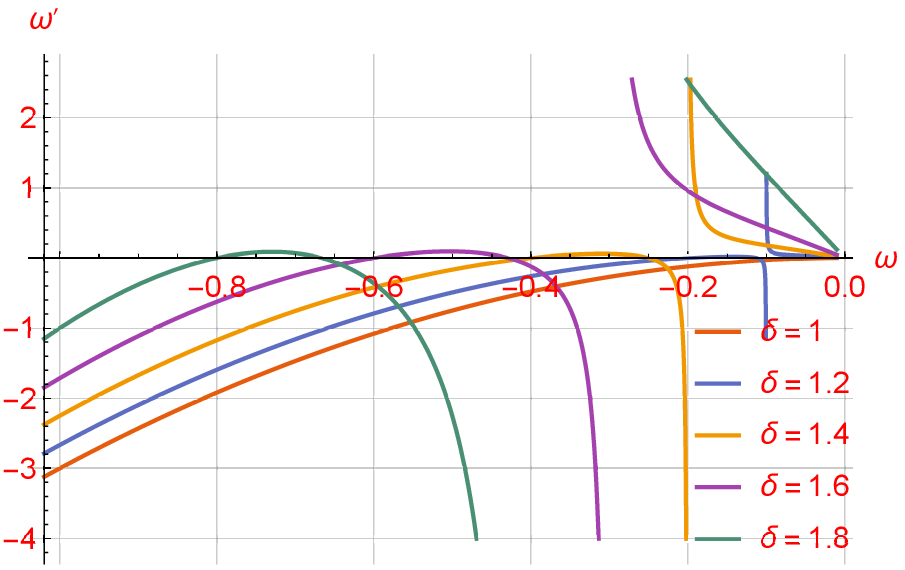}
 \label{1a}}
 \subfigure[]{
 \includegraphics[height=5.5cm,width=5.5cm]{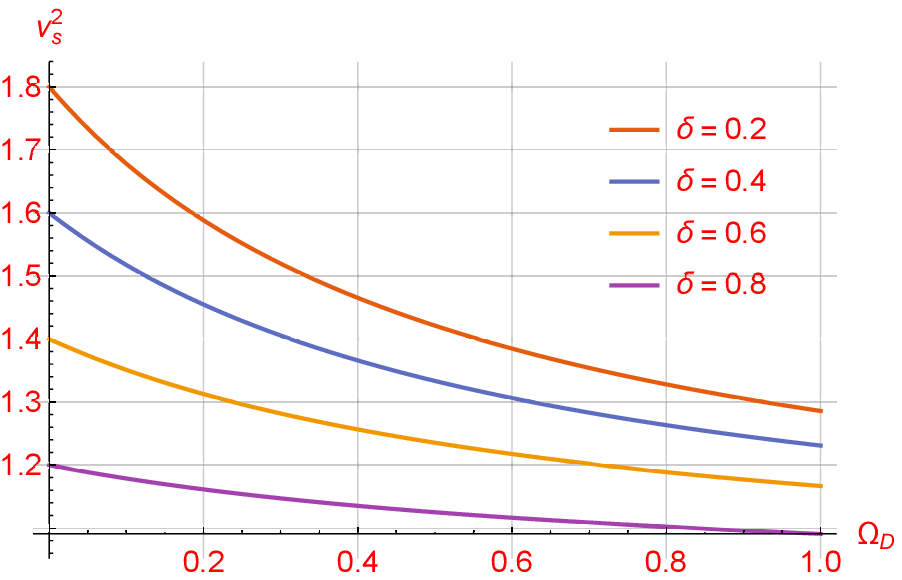}
 \label{1b}}
 \subfigure[]{
 \includegraphics[height=5.5cm,width=5.5cm]{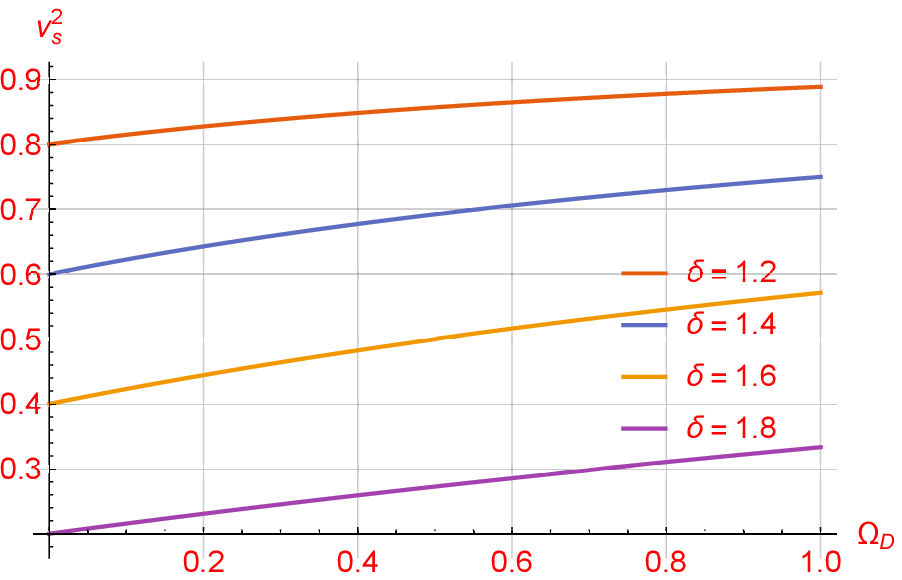}
 \label{1c}}
 \subfigure[]{
 \includegraphics[height=5.5cm,width=5.5cm]{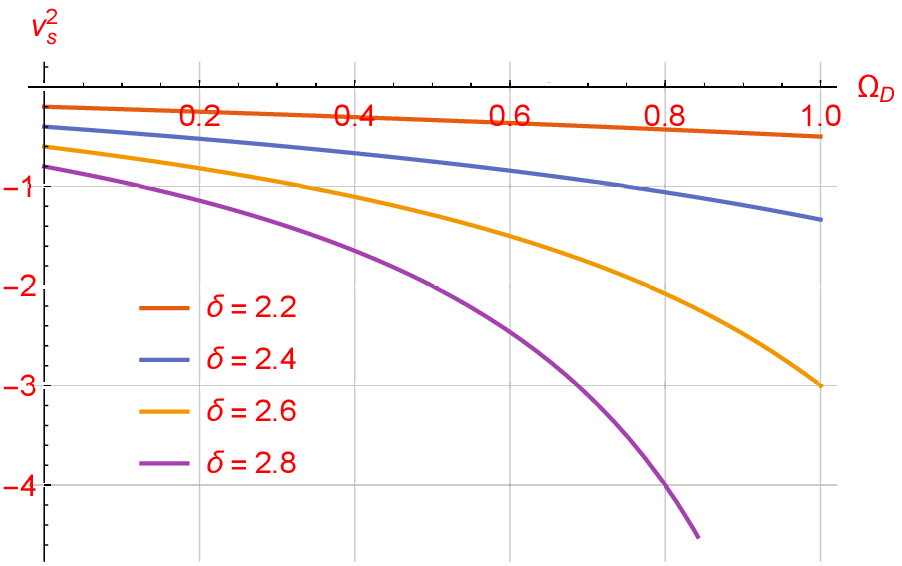}
 \label{1d}}
  \caption{\small{$\omega-\omega'$ analyze in fig (1a), and stability analyze in fig (1b,1c,1d) for the non-interacting case and concerning constant mentioned parameters  }}
 \label{1}
 \end{center}
 \end{figure}

As figure (1a) shows, we can examine the evolutionary trajectories of $\omega-\omega'$  concerning constant parameters $\delta$.
The figure shows that most places offer a negative value concerning constant parameter $\delta$.
Also, for specific values of $\omega$, $\omega'$ equals zero, repeated for all constant parameters.
Of course, this figure also shows that for specific values of $\omega$, the parameter $\omega'$ has a maximum value for each constant parameter. The figure clearly indicates that non-interacting can plays a role in the universe's evolution. Pair analysis for THDE in the non-interacting case is different with ghost dark energy, as discussed in\cite{49}. as shown in the figure for THDE, we face two areas viz part of entirely positive, which can have a minimum in the figure and the other part inverse with it
This figure is plotted for the non-interacting case.
Also, figures (1b,1c,1d) determine the stability in terms of parameter $\Omega_{D}$ for the non-interacting case for the mentioned constant parameters.
As it is clear, corresponding to the constant parameter $\delta$ in the ($0<\delta<2$), the figures take positive values, which indicates the model's stability in the mentioned framework.
 But for the $\delta>2$, the model has negative values in all areas, and it is always in an unstable state.
However, it can be stated that the existence of stability and instability for different values can also be consistent with the previous findings of Tsallis holographic dark energy.
The results align with some of the works and have some differences from someone's\cite{33,34,36,37,42,49}.
Of course, we note that the mentioned model does not assume a stable form for all the assessed values, which can be an influential point because the instability of the model can indicate an important issue. Since the magnitude of the speed of sound cannot be negative, the non-interacting dark energy-dominated universe in the future cannot be expected to be the universe's fate if the model is unstable. This is a general conclusion for this dark energy model, irrespective of whether the complex part of the scalar field is considered or not.
\section{THDE and CQF in FRW universe (interacting case)}

This section will go through a similar process and compare the results obtained. So with respect to equations (16) and (24), we will have,

\begin{equation}\label{47}
\omega_{D}=-\frac{3+(\delta-2)(\frac{k}{H^{2}a^{2}}+3)+3b^{2}(1+(-1+1/\Omega_{D}(1+\Omega_{k}))}{3(1+(\delta-2)\Omega_{D})}.
\end{equation}

With respect to flat universe, i.e., $k=0=\Omega_{k}$

\begin{equation}\label{48}
\omega_{D}=-\frac{3+3(\delta-2)+3b^{2}/\Omega_{D}}{3(1+(\delta-2)\Omega_{D})}.
\end{equation}

For simplicity

\begin{equation}\label{49}
\mathcal{X}=-\omega_{D}=\frac{3+(\delta-2)(\frac{k}{H^{2}a^{2}}+3)+3b^{2}(1+(-1+1/\Omega_{D}(1+\Omega_{k}))}{3(1+(\delta-2)\Omega_{D})}.
\end{equation}

Since we are looking for a correspondence between the energy density of the complex form of quintessence field and Tsallis holographic dark energy, we will have a relation as $\omega_{D}=\omega_{\Phi}$. then

\begin{equation}\label{50}
\frac{\mathcal{T}-V(\phi)}{\mathcal{T}+V(\phi)}=-\mathcal{X}.
\end{equation}

yield

\begin{equation}\label{51}
V(\phi)=-\frac{\mathcal{X}+1}{\mathcal{X}-1}\times\mathcal{T}.
\end{equation}

If we combine equations (26) and (51), we will have

\begin{equation}\label{52}
-\frac{\mathcal{X}+1}{\mathcal{X}-1}\times\mathcal{T}=B H^{4-2\delta}-\mathcal{T}.
\end{equation}

yield

\begin{equation}\label{53}
H=2^{\frac{1}{4-2\delta}}\bigg(\frac{\mathcal{T}}{B-B\mathcal{X}}\bigg).
\end{equation}

concerning equation (49), one can obtain

\begin{equation}\label{54}
H=2^{\frac{1}{4-2\delta}}\bigg(\frac{\mathcal{T}}{B(1+\omega_{D})}\bigg).
\end{equation}

For the relationship to be self-consistent, there must be $\mathcal{X}<1$. To be sure of the universe's accelerated expansion according to equation (7), we can obtain

\begin{equation}\label{55}
\rho_{m}<2\big(V(\phi)-(\dot{\phi}^{2}+\phi^{21}\dot{\theta}^{2})\big)=2V(\phi)-4\mathcal{T}.
\end{equation}

Therefore, it is obtained by using equation (26).

\begin{equation}\label{56}
\rho_{m}<2V(\phi)-4\mathcal{T}=6V(\phi)-4B H^{4-2\delta},
\end{equation}

Thus

\begin{equation}\label{57}
V(\phi)>\frac{2}{3}B H^{4-2\delta}.
\end{equation}

We can set a constraint for the potential concerning equations (26) and (57)

\begin{equation}\label{58}
\frac{2}{3}B H^{4-2\delta}<V(\phi)<B H^{4-2\delta}.
\end{equation}

In continuation, we will advance the ($\omega-\omega'$) and stability analysis for the model in the interacting case, so we have according to equation (24).

\begin{equation}\label{59}
-\frac{3+3(\delta-2)+3b^{2}/\Omega_{D}}{3(1+(\delta-2)\Omega_{D})}.
\end{equation}

Also we will have,

\begin{equation}\label{60}
\frac{\textrm{d}\Omega_{D}}{\textrm{d}\ln a}=\frac{-2+2\delta}{3}\Omega_{D}(1+q),
\end{equation}

\hspace{-0.6cm}where

\begin{equation}\label{61}
q=-\frac{3+(\delta-2)(\Omega_{k}+3)+3b^{2}(1+r))}{3(1+(\delta-2)\Omega_{D})}.
\end{equation}

With respect to above equations, one can obtain,

\begin{equation}\label{62}
\frac{\textrm{d}\Omega_{D}}{\textrm{d}\ln a}=(\delta-1)\Omega_{D}\frac{-3b^{2}(1+\Omega_{k})-3\Omega_{D}+\Omega_{k}+3}{1+(\delta-2)\Omega_{D}}.
\end{equation}

By combining equations (59) and (62), one can calculate,

\begin{equation}\label{63}
\omega'_{D}=\frac{(-1+\delta)\big((-3+\delta)(-2+\delta)\Omega_{D}^{2}+b^{2}(-1-2(-2+\delta)\Omega_{D})\big)(-3+3\Omega_{D}-\Omega_{k}+3b^{2}(1+\Omega_{k})}{\Omega_{D}(1+(-2+\delta)\Omega_{D})^{3}}.
\end{equation}

\begin{figure}[h!]
 \begin{center}
 \subfigure[]{
 \includegraphics[height=5.5cm,width=5.5cm]{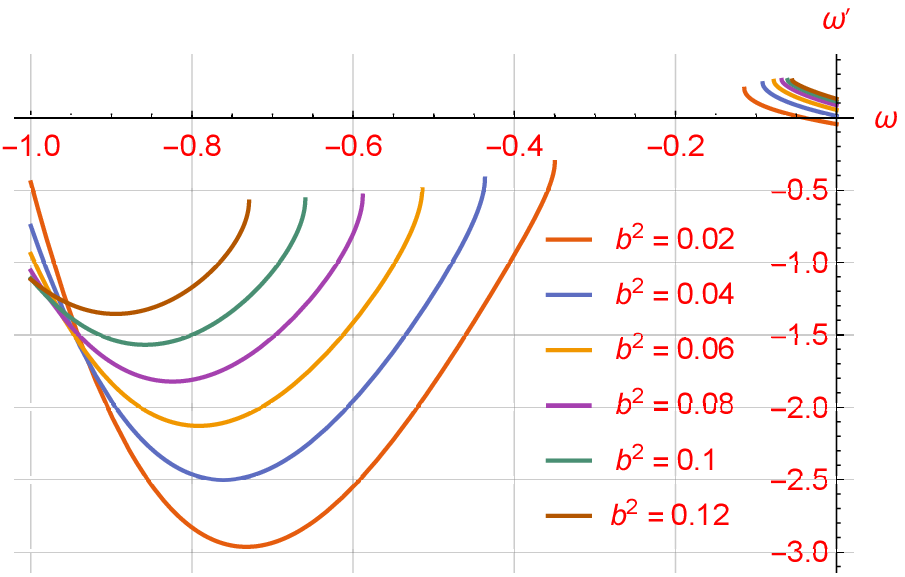}
 \label{2a}}
 \subfigure[]{
 \includegraphics[height=5.5cm,width=5.5cm]{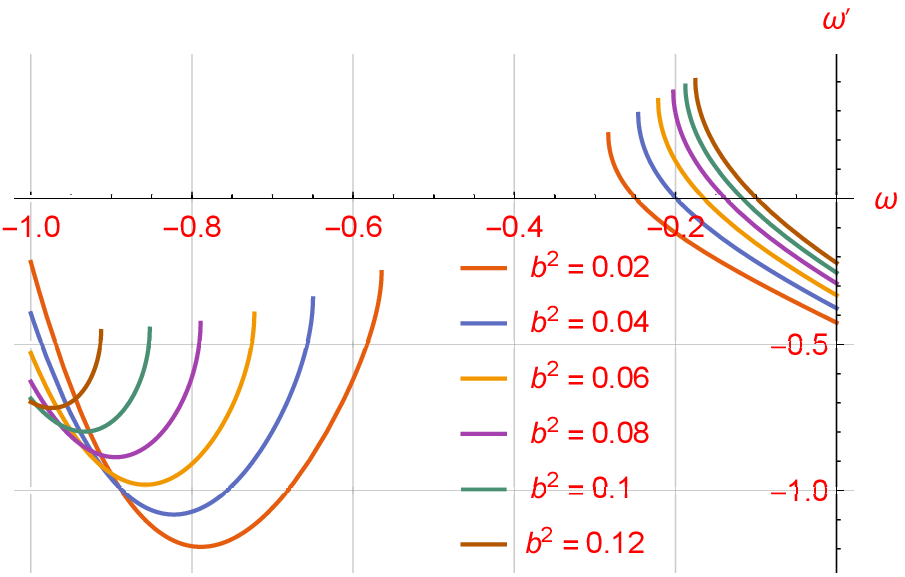}
 \label{2b}}
 \subfigure[]{
 \includegraphics[height=5.5cm,width=5.5cm]{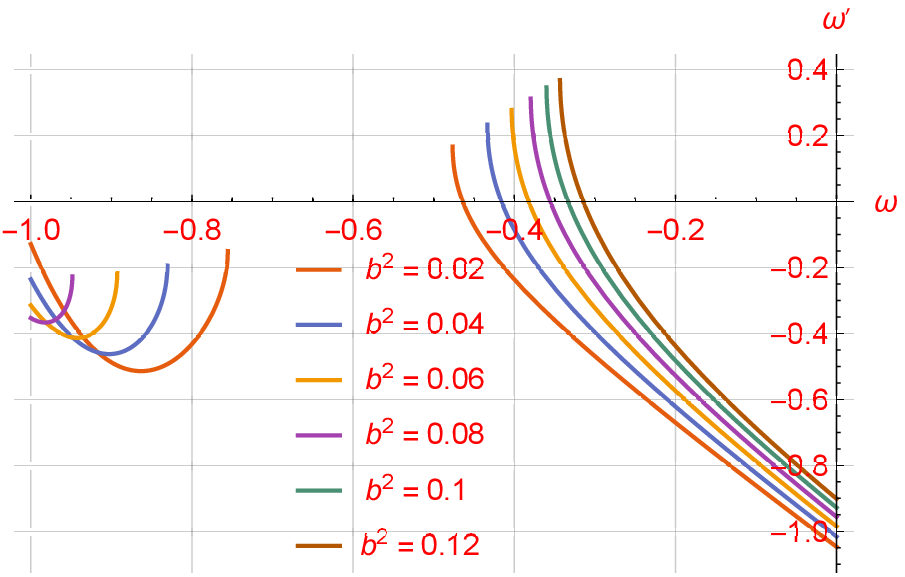}
 \label{2c}}
 \subfigure[]{
 \includegraphics[height=5.5cm,width=5.5cm]{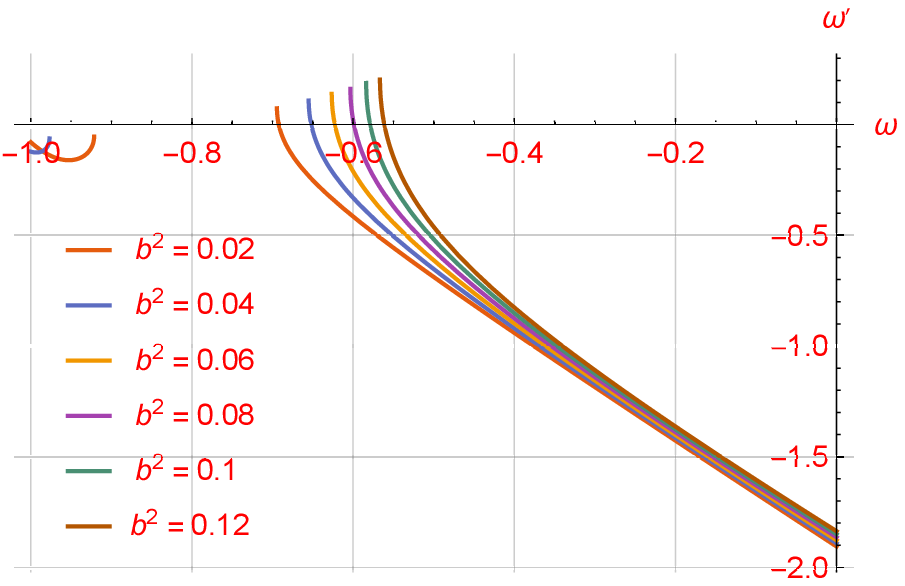}
 \label{2d}}
  \caption{\small{$\omega-\omega'$ analyze for the first sample of interacting case and concerning constant mentioned parameters and $\delta=1.2, 1.4, 1.6, 1.8$ in figs (2a, 2b, 2c, 2d) respectively}}
 \label{2}
 \end{center}
 \end{figure}

Using a similar process, we will examine the stability of this case so that we will have.

\begin{equation}\label{64}
\frac{\textrm{d}H}{\textrm{d } a}=\frac{3^{\frac{1}{2-2\delta}}(3-3\Omega_{D}+\Omega_{k}+3b^{2}(1+\Omega_{k}))(M_{p}/B\Omega_{D})^{\frac{1}{2-2\delta}}}{2a(1+(-2+\delta)\Omega_{D})}.
\end{equation}

So the stability calculated for the flat universe $(k=0=\Omega_{k})$ with respect to equations (37),(40), (41) and (64)

\begin{equation}\label{65}
\nu_{s}^{2}=\frac{(-3+\delta)\Omega_{D}(1-(-2+\delta)^{2}\Omega_{D})+b^{2}(-2+\delta)(1+(-3+2\delta)\Omega_{D})}{\Omega_{D}(1+(-2+\delta)\Omega_{D})^{2}}.
\end{equation}

\begin{figure}[h!]
 \begin{center}
 \subfigure[]{
 \includegraphics[height=4cm,width=4cm]{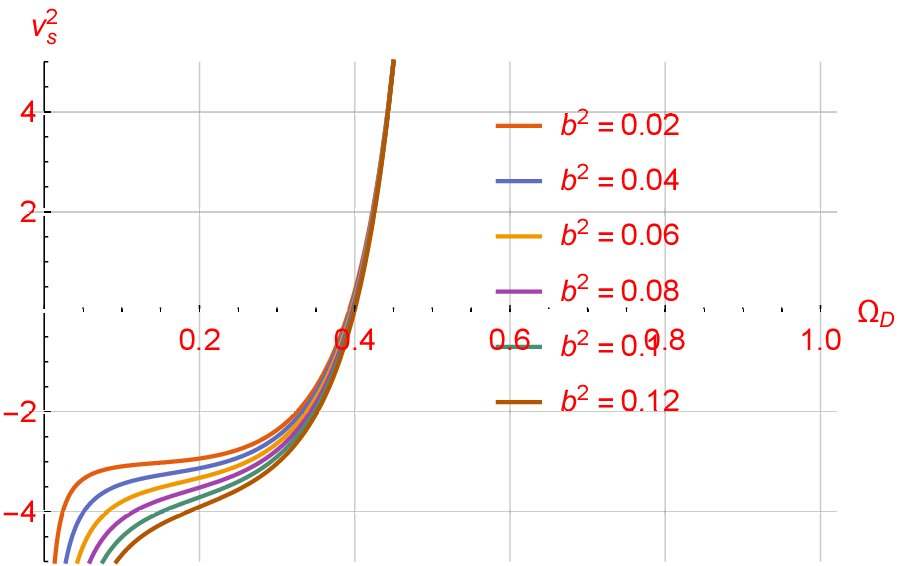}
 \label{3a}}
 \subfigure[]{
 \includegraphics[height=4cm,width=4cm]{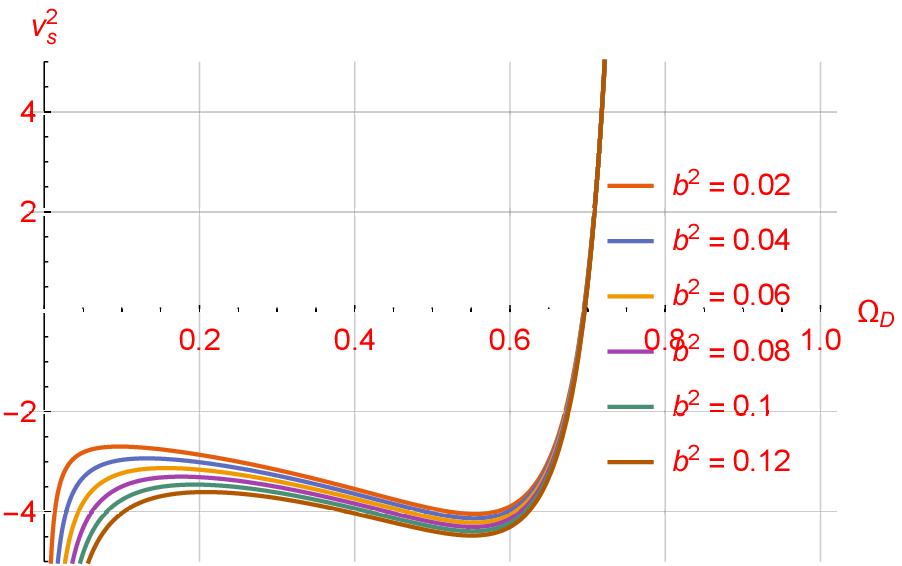}
 \label{3b}}
 \subfigure[]{
 \includegraphics[height=4cm,width=4cm]{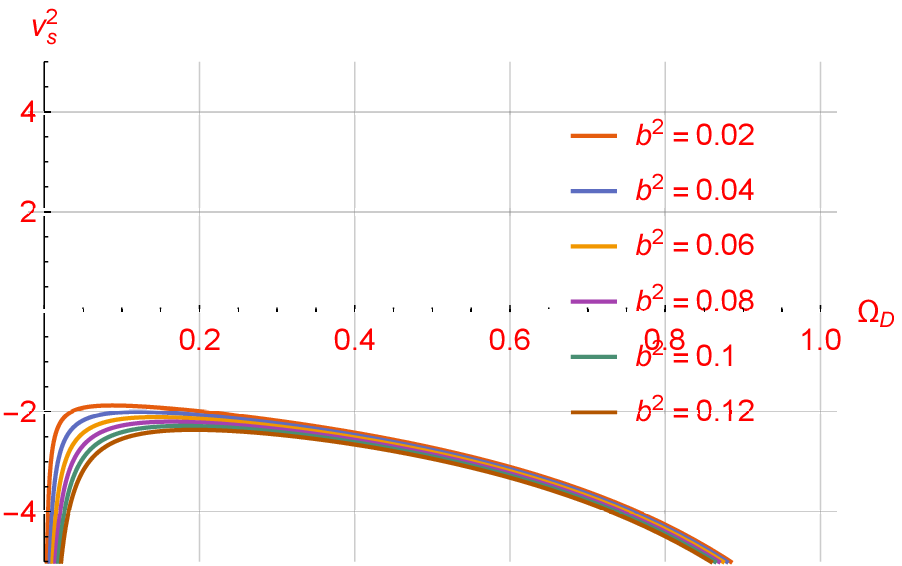}
 \label{3c}}
 \subfigure[]{
 \includegraphics[height=4cm,width=4cm]{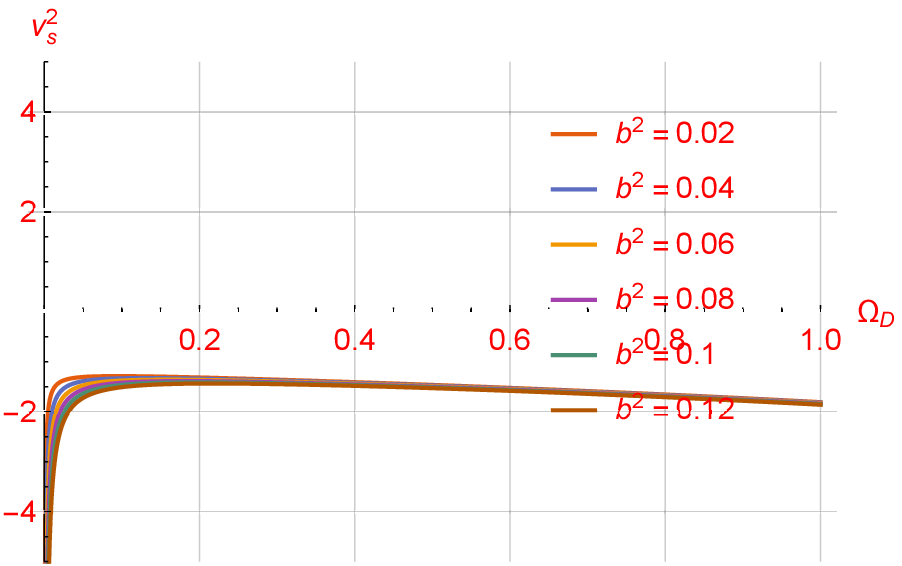}
 \label{3d}}
 \subfigure[]{
 \includegraphics[height=4cm,width=4cm]{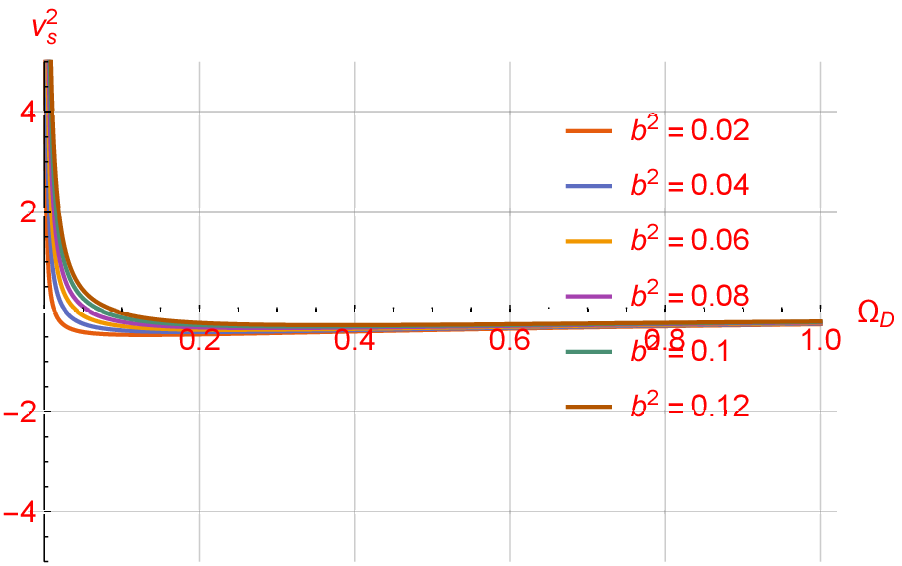}
 \label{3e}}
 \subfigure[]{
 \includegraphics[height=4cm,width=4cm]{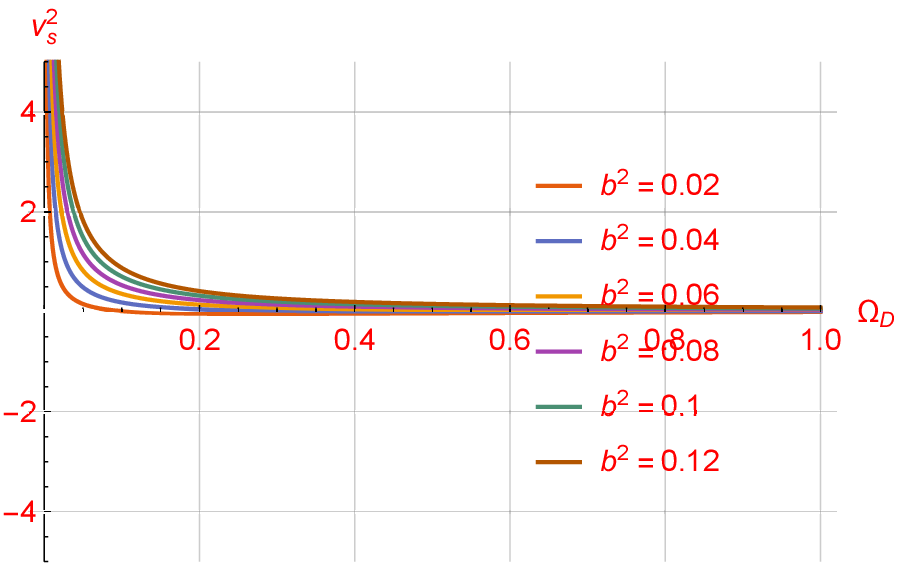}
 \label{3f}}
  \caption{\small{The stability analyze for the interacting case and concerning constant mentioned parameters and $\delta=0.4, 0.8, 1.4, 1.8, 2.4, 2.8$ in figs (3a, 3b, 3c, 3d, 3e,3f) respectively }}
 \label{3}
 \end{center}
 \end{figure}

Figure (2) shows the evolutionary path of $\omega-\omega'$ in exchange for the constant parameters $b^{2}$ and $\delta$ for positive states in this structure.
In this case, unlike the non-interacting, the parameter $b^{2}$ also plays a significant role.
In these calculations, we assumed $\Omega_{D}=0.73$. Like the non-interacting case, we have positive and negative regions for the $\omega-\omega'$ evolutionary path. for some parameter $\omega$ values,  $\omega'$ equals zero. Also, at some points of $\omega$, it has a minimum. From fig 2, we conclude that for the interacting case, there exists some overlapping region in which a value of $\omega$ corresponds to some possible values of $\omega'$. In particular, the width of the region becomes narrower in some areas specified in the figure. This result differs from that in an agegraphic dark energy model and is somehow consistent with Interacting ghost dark energy\cite{49}. The $\omega-\omega'$ analysis is a functional dynamic analysis for discriminating different dark energy models
Figure 3 shows the stability of the model for the interacting case corresponding to the constant parameters $b^{2}$ and $\delta$ in terms of the $\Omega_{D}$. Like the non-interacting case, for this part, the stability of the model is affected by specific values assumed for the parameter $\delta$.
As it is evident in figures (3a) and (3b), for $\delta=0.4$ and $\delta=0.8$ as well as other constant parameters, a part of the figures has positive values in $\Omega_{D}\geq 0.4$ and $\Omega_{D}\geq 0.7$, so it has negative values in other area as well.
The stability of the model is due to the values that make the figures positive, the details of which are fully specified in these two diagrams.
As it is known, there are some differences compared to the non-interacting case. It is related to the negative part, which shows the instability of the model for these constant values, which was not seen in the non-interacting case.
Also, in figures (3c) and (3d), unlike the non-interacting case, the figures take negative values in all area. Hence, it indicates the instability of the model for values of the free parameter $\delta$, i.e., in the range of $1<\delta<2$. As we saw in the previous section, for the non-interacting case, the model was stable for these values.
In figures (3e) and (3f), if we carefully look at the $\delta$ values, the larger this constant parameter, the model will include positive values throughout the region, which means that our model is stable in the mentioned framework.
Contrary to the result obtained for the values of $\delta>2$ in the non-interacting case.
Of course, a similar explanation can be considered for this part about the negative areas that lead to the instability of the mentioned model.
 The model is unstable for different constant parameters $b^{2}$ and $\delta$ if it shows the figures of negative values.
 For the model to be stable, $\nu_{s}^{2}$ must always be a positive values, so, like the non-interacting example, the model still has negative values for specific values in particular regions. That is, it is unstable.
For such negative areas that indicate instability, it cannot be concluded that Tsallis holographic dark energy dominates the universe, and the future is the real universe's destiny.
The results related to the stability of the model in the interacting case for different values of free parameters, like the non-interacting part, have similarities and compatibility results with several other Tsallis holographic dark energy models.
Also, it shows a significant difference with some of these models in other configurations, which you can see for further study\cite{33,34,36,37,42,49}. With a closer look, we notice that the allowable values of free parameters are determined for model stability for our model in the desired framework for both interacting and non-interacting samples. The difference between the interacting and non-interacting cases is specified. It has also been determined in which areas and according to which values of free parameters, and the mentioned model can be a good option for investigating the changes in the universe's fate. Since the magnitude of the speed of sound cannot be negative, an interacting dark energy-dominated universe in the future cannot be expected to be the universe's fate for the unstable cases.

Like the previous section, we will study the effect of the complex part of the quintessence field for this case, i.e., interacting sample. Like the previous section, we will consider only the slow-rolling field effect. So for this model in the interacting case, one can calculate.

\begin{equation}\label{66}
\frac{\omega^{2}}{a^{6}\phi^{2}}=B H^{4-2\delta}\bigg(1-\frac{3+(\delta-2)(\Omega_{k}+3)+3b^{2}/\Omega_{D}}{3(1+(\delta-2)\Omega_{D})}\bigg).
\end{equation}

So we will have,

\begin{equation}\label{67}
\phi=\frac{\omega\Omega_{D}H^{\delta}}{a^{3}H^{2}}\sqrt{\frac{-3-3(-2+\delta)\Omega_{D}}{B(3b^{2}-(-2+\delta)\Omega_{D}^{2}(-3+3\Omega_{D}-\Omega_{k}))}}.
\end{equation}

Like the previous section, we need that $\frac{\textrm{d}\ln \phi}{\textrm{d} a}\approx0$. So

\begin{equation}\label{68}
\begin{split}
&0\approx \bigg\{3(-1+\delta)\Omega_{D}(-1+b^{2}+\Omega_{D})((-2+\delta)(-1+\delta)\Omega_{D}^{3}\\
&+b^{2}(2+3(-2+\delta)\Omega_{D}))\omega H^{\delta}\bigg\}\bigg/\bigg\{2(-1-(-2+\delta)\Omega_{D})^{3/2}\\
&\times(-b^{2}+(-2+\delta)(-1+\Omega_{D})\Omega_{D}^{2})\sqrt{a^{6}B(b^{2}-(-2+\delta)(-1+\Omega_{D})\Omega_{D}^{2})H^{4}}\bigg\}.
\end{split}
\end{equation}
In this article, parameters $\delta$ and $b^{2}$ play a vital role in calculations; Unlike other works, here we established a relationship between several parameters so that we can use these relations to set limits on their upper and lower bounds. According to the calculations, the parameter $b^{2}$ limitations for our solutions to be acceptable are specified in the below diagram. We have considered ($\Omega_{D}=0.73$). Therefore, according to the above equation, for different values of the parameter($\Omega_{D}$) and ($\delta$), various values are obtained for $b^{2}$. In fact, $b^{2}$ are related to parameter ($\Omega_{D}$) and ($\delta$). For this reason, we plot figure (4).
 This figure shows the changes in these two parameters about each other.
In different calculations,  various results have been obtained for the parameter ($b^{2}$), including (0.08), (0.09), etc,.
The $b^{2}$ in our calculations is also in line with the results obtained in the literature.
$b^{2}$ were considered a free parameter in many calculations.
In this article, we used the effect of the complex part of the quintessence field. As it is apparent in the final equation, the $\Omega_{D}$ can not be any arbitrary value between 0 and 1,i.e.,  the selection of this parameter must guarantee the allowable values for the parameter $(b^{2})$.
\begin{figure}[h!]
 \begin{center}
 \subfigure[]{
 \includegraphics[height=5cm,width=5cm]{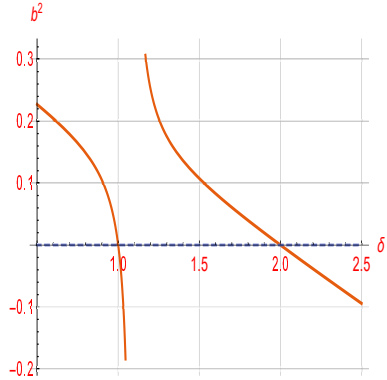}
 \label{4a}}
  \caption{\small{The changes of component $b^{2}$ in terms of $\delta$ }}
 \label{4}
 \end{center}
 \end{figure}

\section{Conclusion}
In this paper, we used a Tsallis holographic dark energy model in two forms, interacting and non-interacting cases, to acquire some parameters as the equation of state for the energy density of the Tsallis model in the FRW universe concerning the complex form of the quintessence model. We studied the cosmology of complex quintessence by revamping the potential and investigating the scalar field dynamics. Then we analyzed ($\omega-\omega'$) and stability in two cases, i.e., non-interacting and interacting. We explored whether these cases describe a real universe by calculating fractional energy density $\Omega_{D}$ and concerning two parts of the quintessence field effect ( complex and real part ) by considering the real part of this field to be a slow-roll field.
We know that the part in which the fractional energy density ($\Omega_{D} > 1$) does not describe a real universe. Also, we specified an interacting coupling parameter $b^{2}$ that depends on the constant parameter of the Tsallis holographic model ($\delta$) with respect to fractional energy density ($0.73$). Unlike independence between the fractional energy density and interacting coupling in the real quintessence model, we determined a relationship among these parameters in this theory. Finally, by plotting some figures, we specified the features of ($\omega-\omega'$) and ($\nu_{s}^{2}$) in two cases and compared the result with each other. We showed that in the non-interacting case, the evolution path of $\omega-\omega'$ is negative in most points. For specific values of the $\omega$, the parameter $\omega'$ was equal to zero and had maximum points. Also, the system is unstable at all points in this case, as this feature has been checked for Tsallis holographic dark energy.
Also, in the interacting case, the evolution path $\omega-\omega'$ had common points with the non-interacting mode, and in this case, the system was unstable in all places.
Of course, in the interacting case, in addition to $(\delta)$, the parameter $b^{2}$ also played an important role, so that in the end, we specified that $\Omega_{D}$ could not be any arbitrary value between 0 and 1. Still, the two parameters $\Omega_{D}$ and $(\delta)$ must be set so that $b^{2}$ be within their allowable range.\\
Here is an important point that can be raised as an issue for the future. For example, one can investigate the problem studied in this paper for other models of dark energy and holographic dark energy as Kaniadakis's holographic dark energy. Can compare The results with the results of this paper. Second, other models of complex form are calculated, and their correspondence with different dark energy models is studied. Third, this study can be examined in combination with other conditions.\\

Data availability statement\\
There are no new data associated with this work.

\end{document}